\documentclass[10pt,amsmath,amssymb,
				reprint,a4paper,prd,aps,
				showpacs,showkeys,
				superscriptaddress]{revtex4-1}
\usepackage[english]{babel}
\usepackage[utf8]{inputenc}

\usepackage{MnSymbol}
\usepackage{mathrsfs}
\usepackage[makeroom]{cancel}

\usepackage{graphicx}
\usepackage{subfig}

\usepackage{amsthm}
\newtheoremstyle{CMPstyle}
  {\topsep} 
  {\topsep} 
  {\itshape} 
  {} 
  {\bfseries} 
  {.} 
  {.5em} 
  {} 
\theoremstyle{CMPstyle}

\usepackage[colorlinks=true,
			citecolor=green,
			linkcolor=red,
			filecolor=cyan,
			urlcolor=magenta]{hyperref}

\newcommand{\dd}{\mbox{d}}

\begin{document}
\title{Kerr principal null directions from Bel--Robinson and Kummer surfaces}

\author{Jens Boos}
\email[Corresponding author, E-Mail:~]{boos@ualberta.ca}
\affiliation{Theoretical Physics Institute, University of Alberta, Edmonton, AB T6G 2E1, Canada}
\author{Alberto Favaro}
\email[E-Mail:~]{a.favaro@imperial.ac.uk}
\affiliation{Department of Physics, Imperial College London, UK, and ASI Data Science, London, UK}

\date{\today}


\begin{abstract}
In the Kerr geometry, we calculate various surfaces of constant curvature invariants. These extend well beyond the Kerr horizon, and we argue that they might be of observational significance in connection with non-minimally coupled matter fields. Moreover, we demonstrate that the principal null directions of the Kerr geometry can be obtained by projections involving either the Bel--Robinson or the Kummer tensor. We conjecture that this is also possible in more general settings.

Essay written for the Gravity Research Foundation 2017 Awards for Essays on Gravitation.

{\small \textit{file: 28\_kerr\_v2.tex, Mar 31, 2017, jb} }
\end{abstract}

\maketitle


The Kerr solution \cite{Kerr:1963ud}, describing a rotating black hole, is a truly remarkable exact solution of the vacuum Einstein field equations: it is not only important in astrophysics, has not only spawned a wealth of mathematical tools and devices in the context of the finding of solutions to the Einstein field equations \cite{Kerr:2007dk,Heinicke:2015iva}, but it has also served as a testing ground for new physical ideas. For example, the concept of hidden symmetries \cite{Frolov:2011} was discovered after studying the separability of the geodesic equation and the appearance of Carter's constant \cite{Carter:1968}. More recently, the emergence of conformal symmetry in the near-horizon region of a maximally rotating Kerr black hole has given rise to the Kerr/CFT correspondence \cite{Castro:2010fd}.

In this essay, we would like to focus on properties of the Kerr black hole that can be described using invariant expressions obtained from the curvature. We will apply two techniques: visual analysis of invariant curvature surfaces, and projective methods involving the principal null directions of the Kerr spacetime as contracted with tensorial expressions in the curvature. See also Refs.~\cite{Nichols:2011pu,Abdelqader:2013bma,Koehler} regarding different visualization procedures of spacetime curvature.

The Kerr solution in Boyer--Lindquist coordinates $\{t,r,\theta,\phi\}$ takes the form
\begin{align}
\begin{split}
\dd s^2 &= -\,\frac{\Delta}{\rho^2}\left(\dd t - a \sin^2\theta\,\dd\phi\right)^2 + \frac{\rho^2}{\Delta}\, \dd r^2 + \rho^2\,\dd\theta^2\\
 &\hspace{10pt}+ \frac{\sin^2\theta}{\rho^2}\left[ (r^2 + a^2)\,\dd\phi - a \, \dd t \right]^2 , \label{eq:kerr-metric}
 \end{split}
\end{align}
where we defined $\rho^2 := r^2 + a^2\cos^2\theta$ and $\Delta := r^2 - 2mr + a^2$. We may readily read off a possible orthonormal coframe such that $\dd s^2 = -\vartheta{}^{\hat{0}}\otimes\vartheta{}^{\hat{0}}+\vartheta{}^{\hat{1}}\otimes\vartheta{}^{\hat{1}}+\vartheta{}^{\hat{2}}\otimes\vartheta{}^{\hat{2}}+\vartheta{}^{\hat{3}}\otimes\vartheta{}^{\hat{3}}$.

\section{Invariant curvature surfaces}
The Kerr solution \eqref{eq:kerr-metric} is a vacuum solution, and thus the Riemann curvature tensor equals the Weyl tensor, $R{}_{ijkl} = C{}_{ijkl}$. All invariant and tensorial expressions in curvature shall be derived from the Weyl tensor alone.

Let us start by defining the two Cartan invariants \cite{MacCallum:2015zaa}
\begin{align}
\begin{split}
\mathbb{E} &:= - \, \frac 12 C{}_{\hat{0}\hat{1}\hat{0}\hat{1}} = mr \, \frac{r^2 - 3a^2\cos^2\theta}{(r^2 + a^2\cos^2\theta)^3} , \\
\mathbb{B} &:= \frac 12 C{}_{\hat{0}\hat{1}\hat{2}\hat{3}} = ma\cos\theta \, \frac{3r^2 - a^2\cos^2\theta}{(r^2 + a^2\cos^2\theta)^3} ,
\end{split}
\end{align}
where hatted indices refer to the orthonormal coframe $\vartheta{}^\mu$. These quantities capture the mass (``gravitational charge'') and angular momentum (``gravitational current'') of the Kerr solution. They are ubiquitous in analytic expressions for different (pseudo-)scalar invariants.

\begin{figure*}[!htb]%
    \centering
    \subfloat[Kretschmann scalar $K$]{{\includegraphics[width=0.3\textwidth]{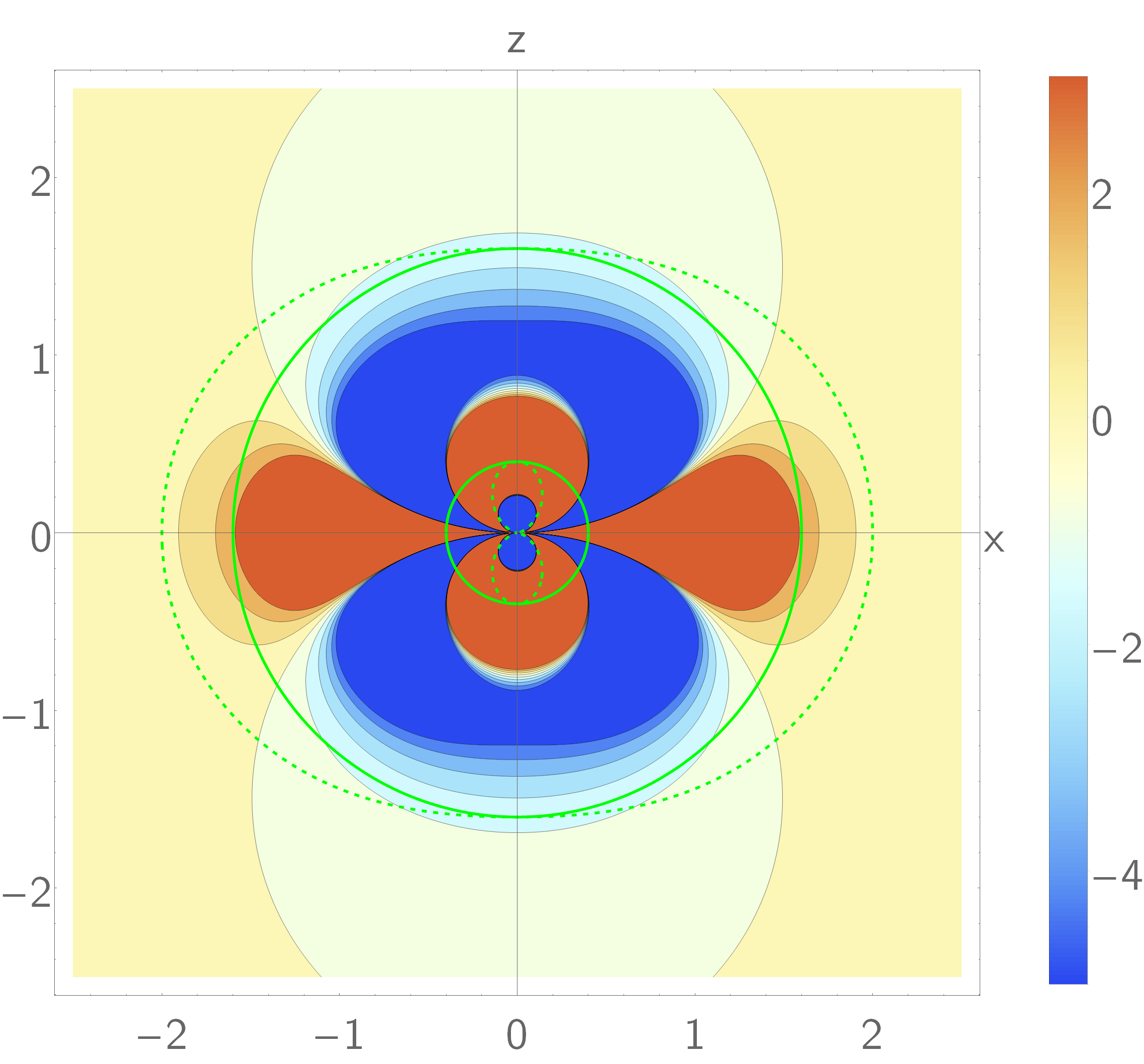} }}%
    \qquad
    \subfloat[Kummer scalar $S$]{{\includegraphics[width=0.3\textwidth]{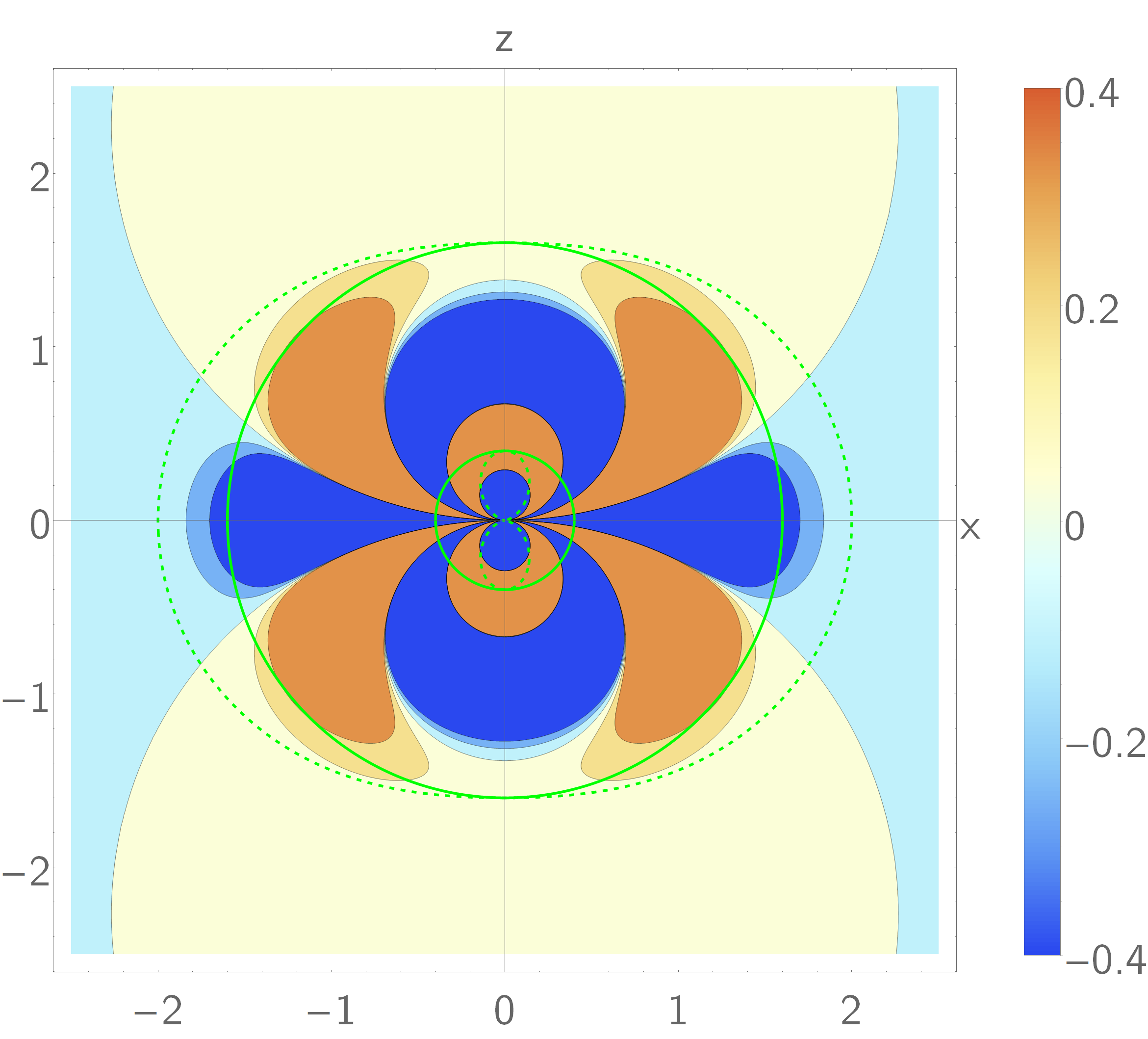} }}%
    \qquad
    \subfloat[Karlhede--Lindstr\"om--\AA man scalar $K'$]{{\includegraphics[width=0.3\textwidth]{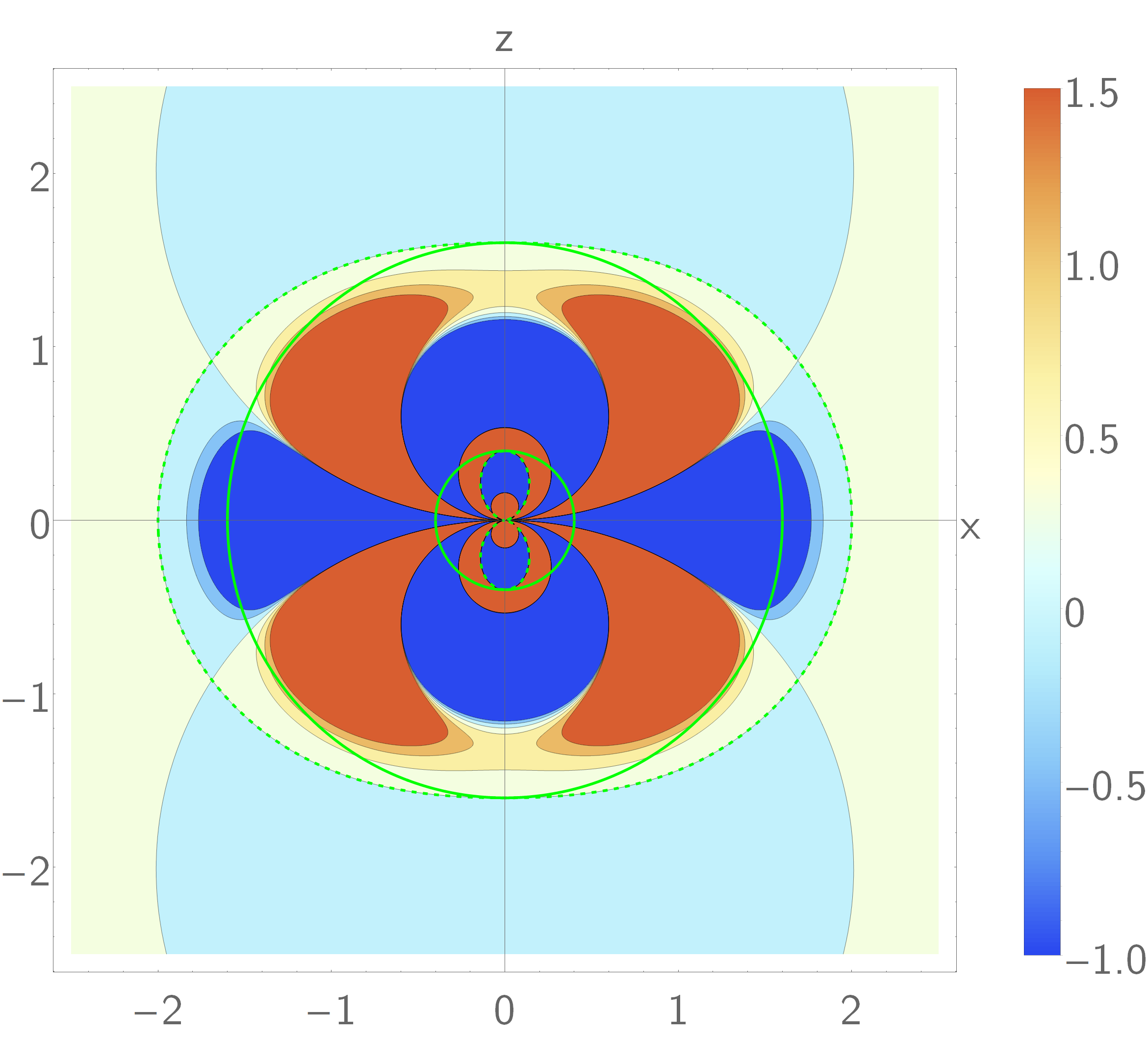} }}
    
    \subfloat[Pontryagin pseudoscalar $\mathcal{P}$]{{\includegraphics[width=0.3\textwidth]{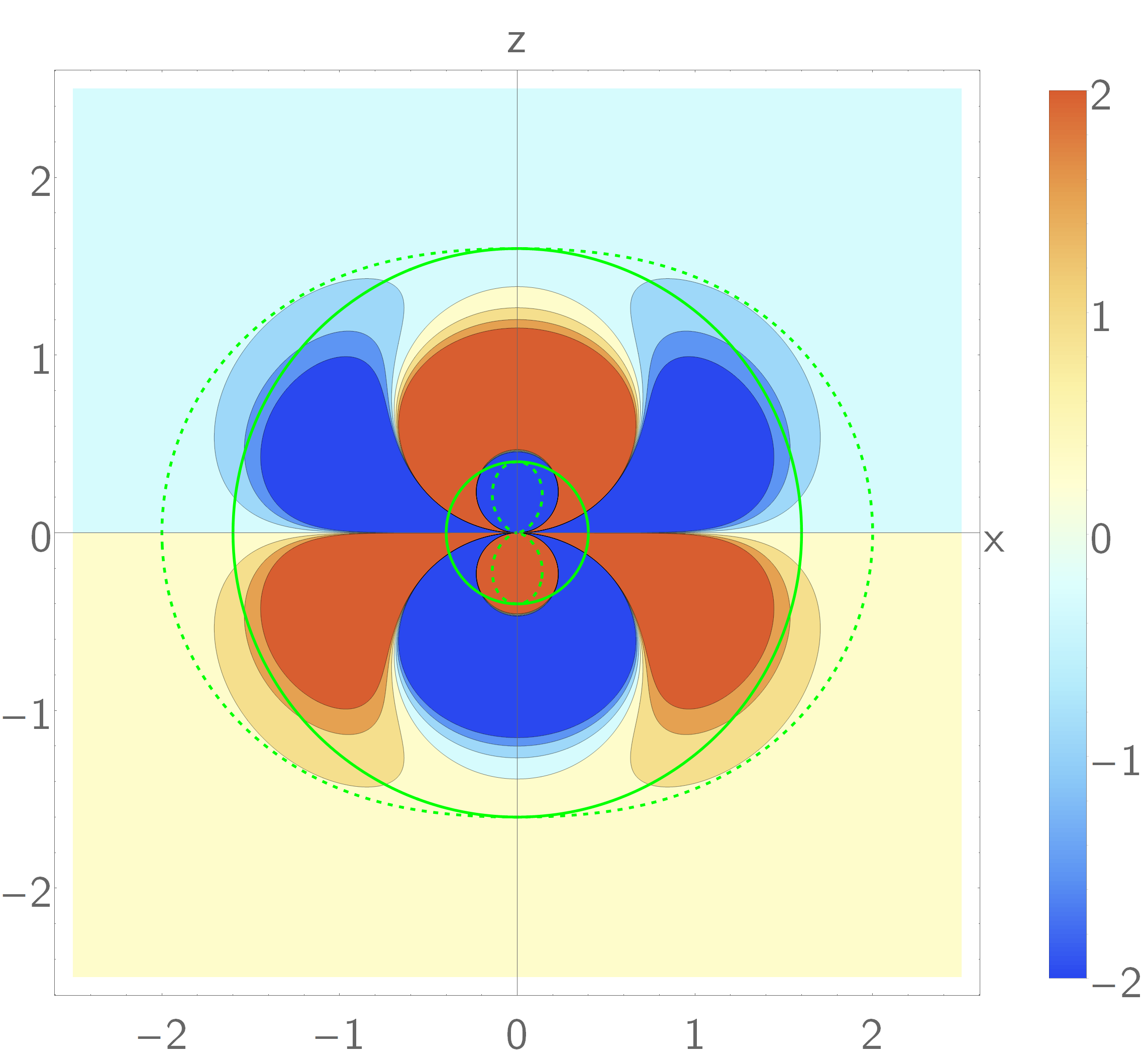} }}%
    \qquad
    \subfloat[Kummer pseudoscalar $\mathcal{A}$]{{\includegraphics[width=0.3\textwidth]{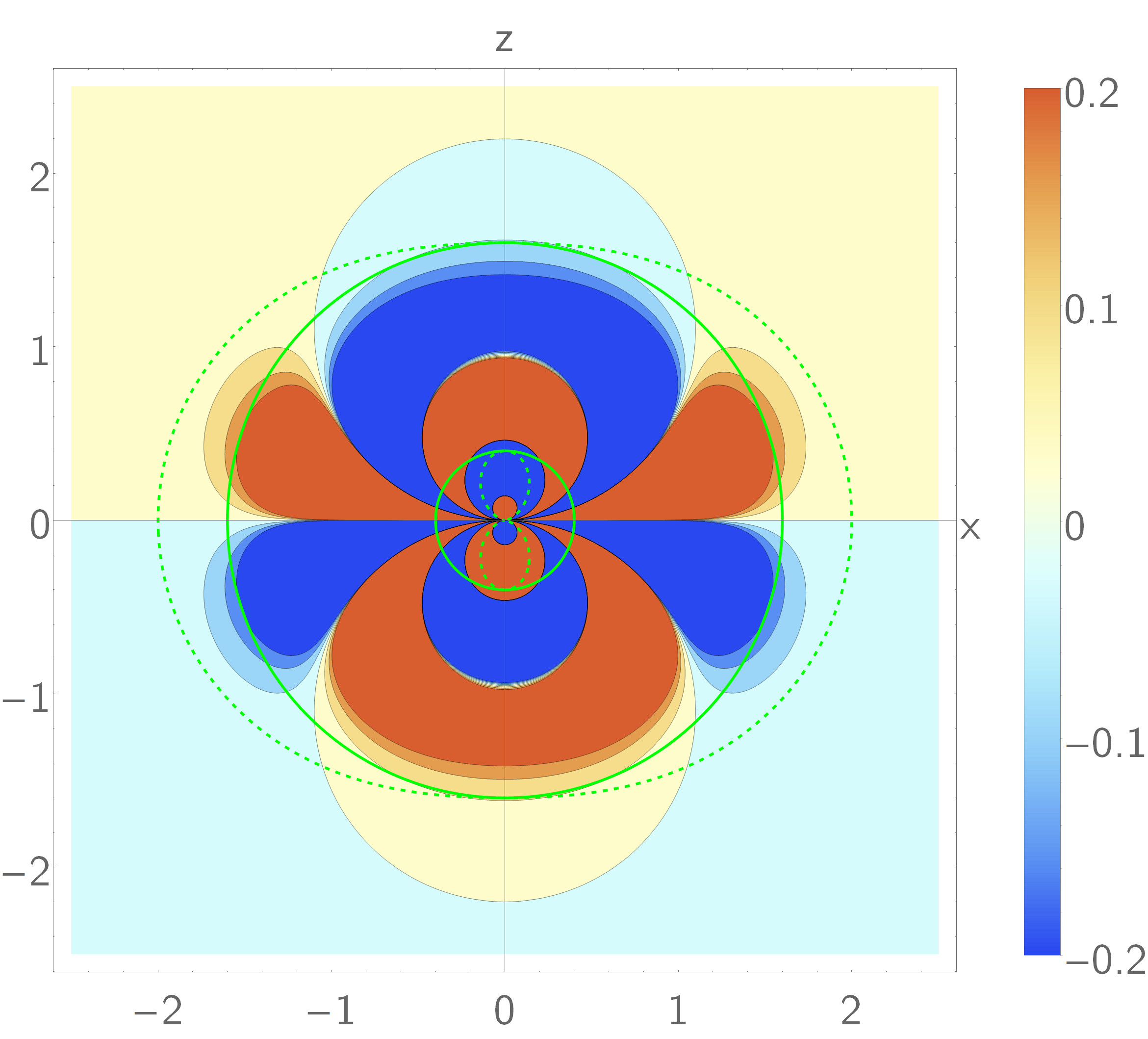} }}%
    \qquad
    \subfloat[Karlhede--Lindstr\"om--\AA man pseudoscalar $\mathcal{P}'$]{{\includegraphics[width=0.3\textwidth]{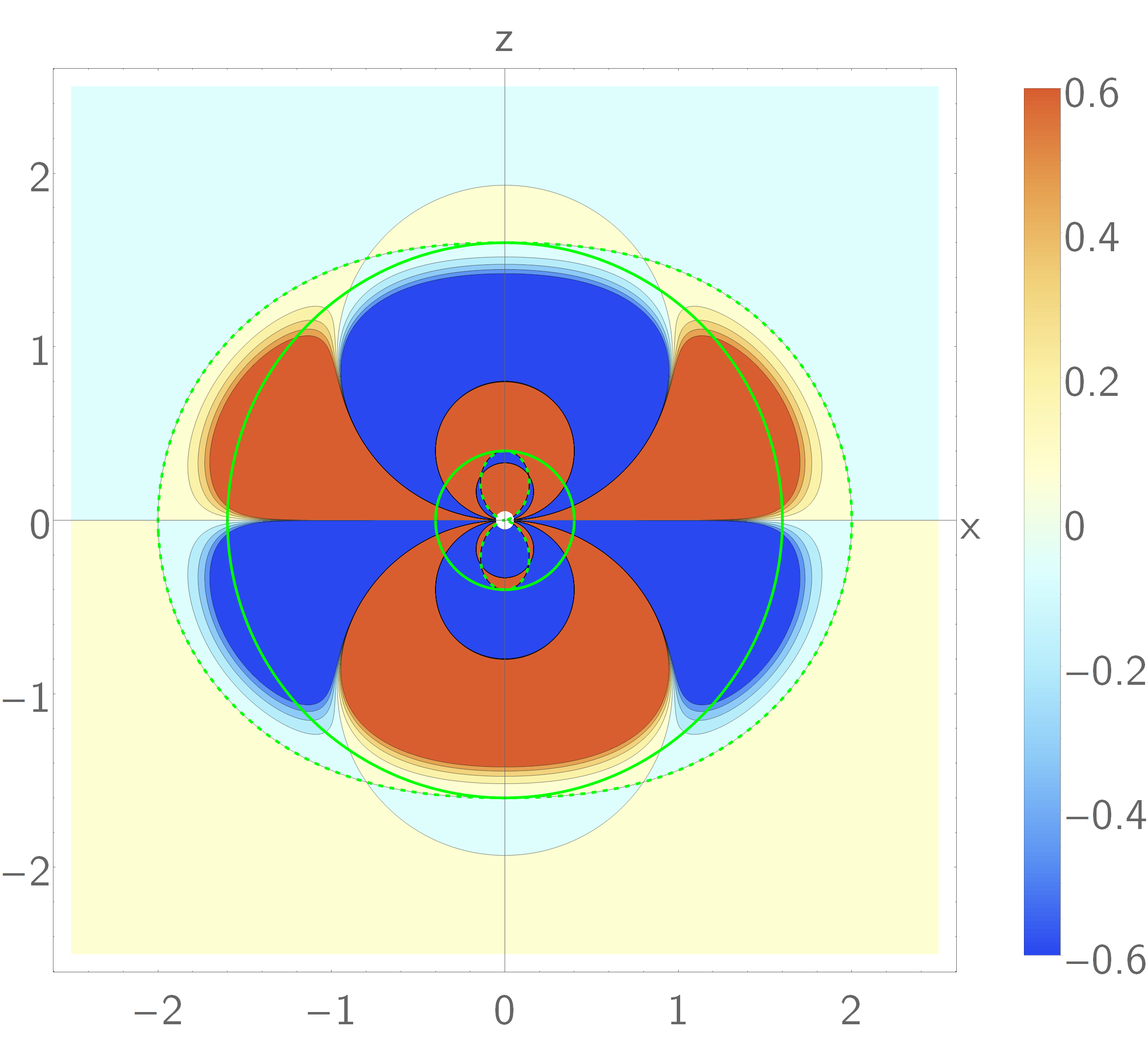} }}%
    \caption{(Colors online.) Various invariants for the Kerr black hole with $m=1$, $a=0.8$, plotted in the plane $y = 0$. The $z$ axis is the axis of rotation, and $r^2 =: x^2+y^2+z^2$. Note that all scalars are reflection symmetric with respect to $z \rightarrow -z$, whereas all pseudoscalars are antisymmetric. The inner and outer ergosphere are indicated by dashed green, whereas the inner and outer horizon correspond to the solid green lines.}%
    \label{fig:2d-invariants}%
\end{figure*}

Focusing the discussion somewhat, let us now introduce the curvature invariants that will be examined in this work. The Kretschmann scalar $K$ and Chern--Pontryagin pseudoscalar $\mathcal{P}$ take the values \cite{Boos:2014hua}
\begin{align}
\begin{split}
K &:= C{}_{abcd}C{}^{abcd} = 48\left( \mathbb{E}^2 - \mathbb{B}^2 \right) , \\
\mathcal{P} &:= \left(\ast C\right){}_{abcd} C{}^{abcd} = -96 \mathbb{E} \mathbb{B} ,
\end{split}
\end{align}
where $\left(\ast C\right){}_{ijkl}$ is the left dual of the Weyl tensor. Following Ref.~\cite{Abdelqader:2014vaa} we consider the additional Karlhede--Lindstr\"om--\AA man invariants \cite{Karlhede:1982}
\begin{align}
\begin{split}
K' &:= \left( \nabla{}_a C{}_{bcde} \right)\left( \nabla{}^a C{}^{bcde} \right), \\
\mathcal{P}' &:= \left[ \nabla{}_a \left(\ast C\right){}_{bcde} \right]\left( \nabla{}^a C{}^{bcde} \right) .
\end{split}
\end{align}
They vanish on the Kerr ergosphere (and hence at the Schwarzschild horizon, for $a=0$), but they are not very useful in locating the horizon of the Kerr solution. Their explicit form can be found in Ref.~\cite{Abdelqader:2014vaa}. By contrast, the following invariant vanishes on the Kerr horizon \cite{Abdelqader:2014vaa,Page:2015aia}:
\begin{align}
\begin{split}
H &:= \star \left[ \dd K \wedge \dd \mathcal{P} \wedge \star \left( \dd K \wedge \dd \mathcal{P} \right) \right] \\
  &= 16\times12^8\times \frac{m^8a^2\cos^2\theta(r^2-2mr+a^2)}{\left(r^2+a^2\cos^2\theta\right){}^{16}} .
\end{split}
\end{align}
We close by defining the Bel--Robinson tensor \cite{Bel:1962,Robinson:1997} as well as the vacuum Kummer tensor \cite{Baekler:2014kha,Baekler:2014b}:
\begin{align}
\begin{split}
B{}_{ijkl} &:= C{}_{iabk} C{}_j{}^{ab}{}_l + \left(\ast C\right){}_{iabk} \left(\ast C\right){}_j{}^{ab}{}_l , \\
K{}_{ijkl} &:= - \, C{}_{iajb} C{}^{acbd} C{}_{kcld} .
\end{split}
\end{align}
The Bel--Robinson tensor is related to the notion of superenergy, since its full contraction with any timelike vector is positive, $B{}_{abcd}u{}^a u{}^b u{}^c u{}^d \ge 0$ \cite{Senovilla:1999xz}. The Kummer tensor can be introduced by analogy with electromagnetism. In Ref.~\cite{Baekler:2014kha}, it was suggested that this tensor may encode specific properties of gravitational waves. The Bel--Robinson tensor and the Kummer tensor admit the following invariants:
\begin{align}
\begin{split}
B &:= B{}_{abcd} B{}^{abcd} = 4 \times 12^2 \times \left( \mathbb{E}^2 + \mathbb{B}^2 \right)^2, \\
S &:= K{}^{ab}{}_{ab} = 48 \mathbb{E}\left( 3\mathbb{B}^2 - \mathbb{E}^2 \right) , \\
\mathcal{A} &:= K{}_{abcd}\epsilon{}^{abcd} = 24 \mathbb{B}\left( 3\mathbb{E}^2 - \mathbb{B}^2 \right) .
\end{split}
\end{align}
Now we can define invariant curvature surfaces by setting
\begin{align}
f(K, \mathcal{P}, K', \mathcal{P}', H, B, S, \mathcal{A}) = \text{const} , \label{eq:invariant-curvature-surface}
\end{align}
where $f$ is some polynomial function. Of course, one may also consider simpler cases where just one of the invariants assumes a constant value, as can be seen in Fig.~\ref{fig:2d-invariants}. These invariants define surfaces that may extend well outside the horizon and can take relatively complicated shapes.

One may ask: what is a special value for these invariants to take? Since the Kerr solution can be written as $\dd s{}^2 = m^2 \dd \tilde{s}^2$, where the new line element $\dd\tilde{s}^2$ only depends on the dimensionless parameter $\alpha := a/m$, there is no intrinsic length scale other than $m$. In other words, if all distances are measured in terms of $m$, the dimensionless parameter $\alpha$ no longer provides a length scale related to the rotation parameter.

\begin{figure*}[!hbt]
    \centering
    \subfloat[$m=1$, $a=0.3$]{{\includegraphics[width=0.38\textwidth]{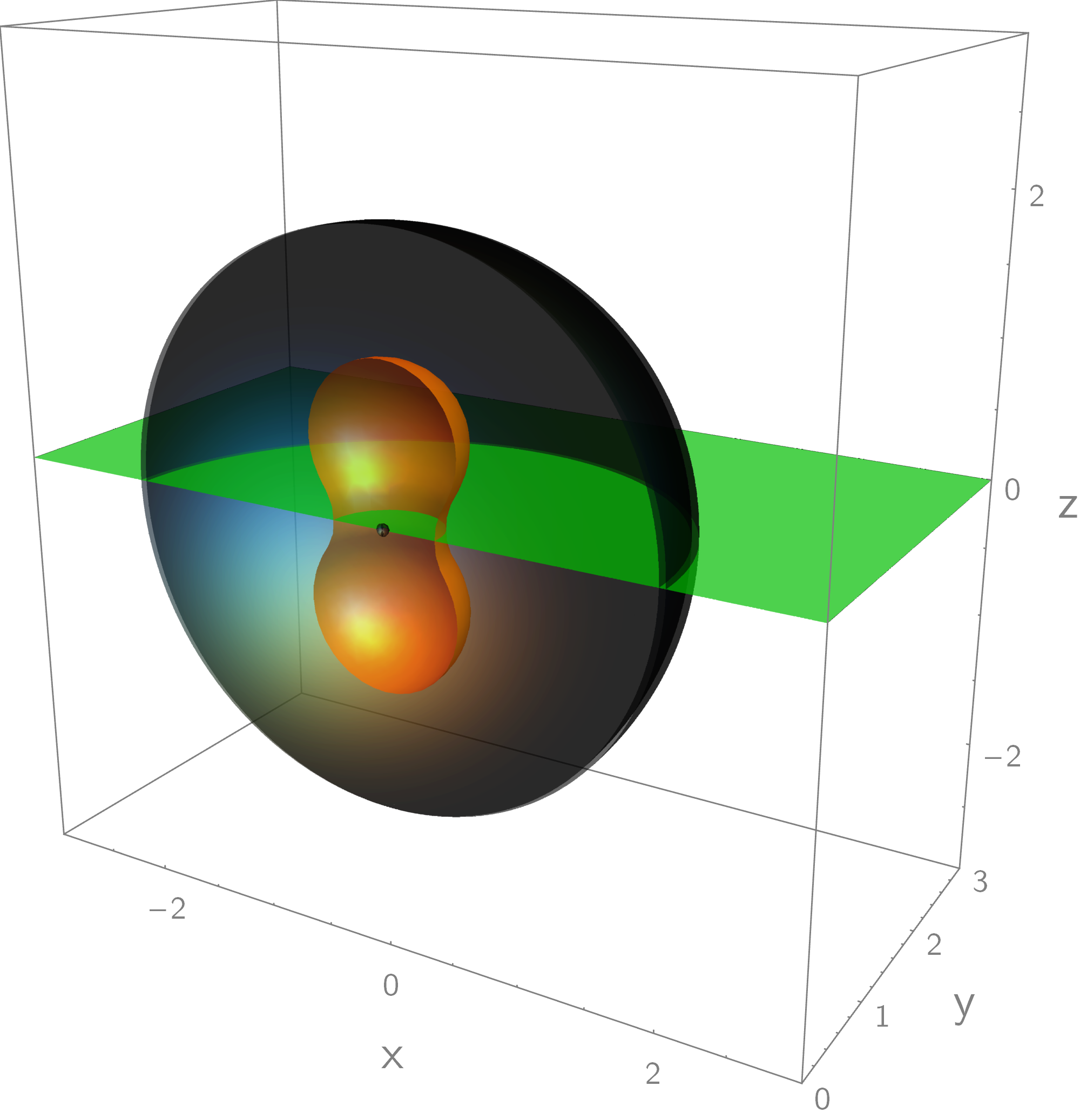} }}%
    \qquad
    \subfloat[$m=1$, $a=0.5$]{{\includegraphics[width=0.38\textwidth]{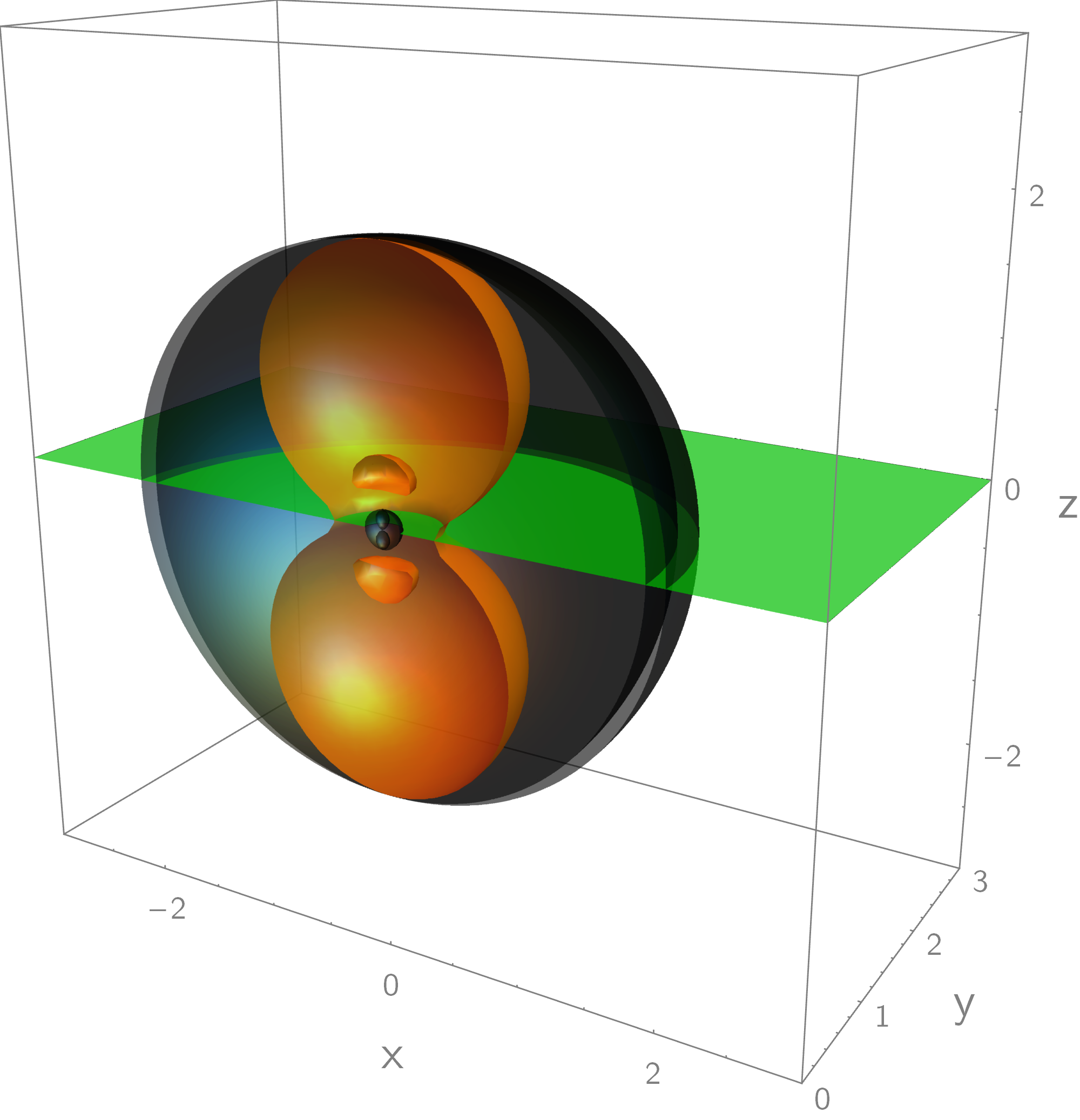} }}%

    \subfloat[$m=1$, $a=0.8$]{{\includegraphics[width=0.38\textwidth]{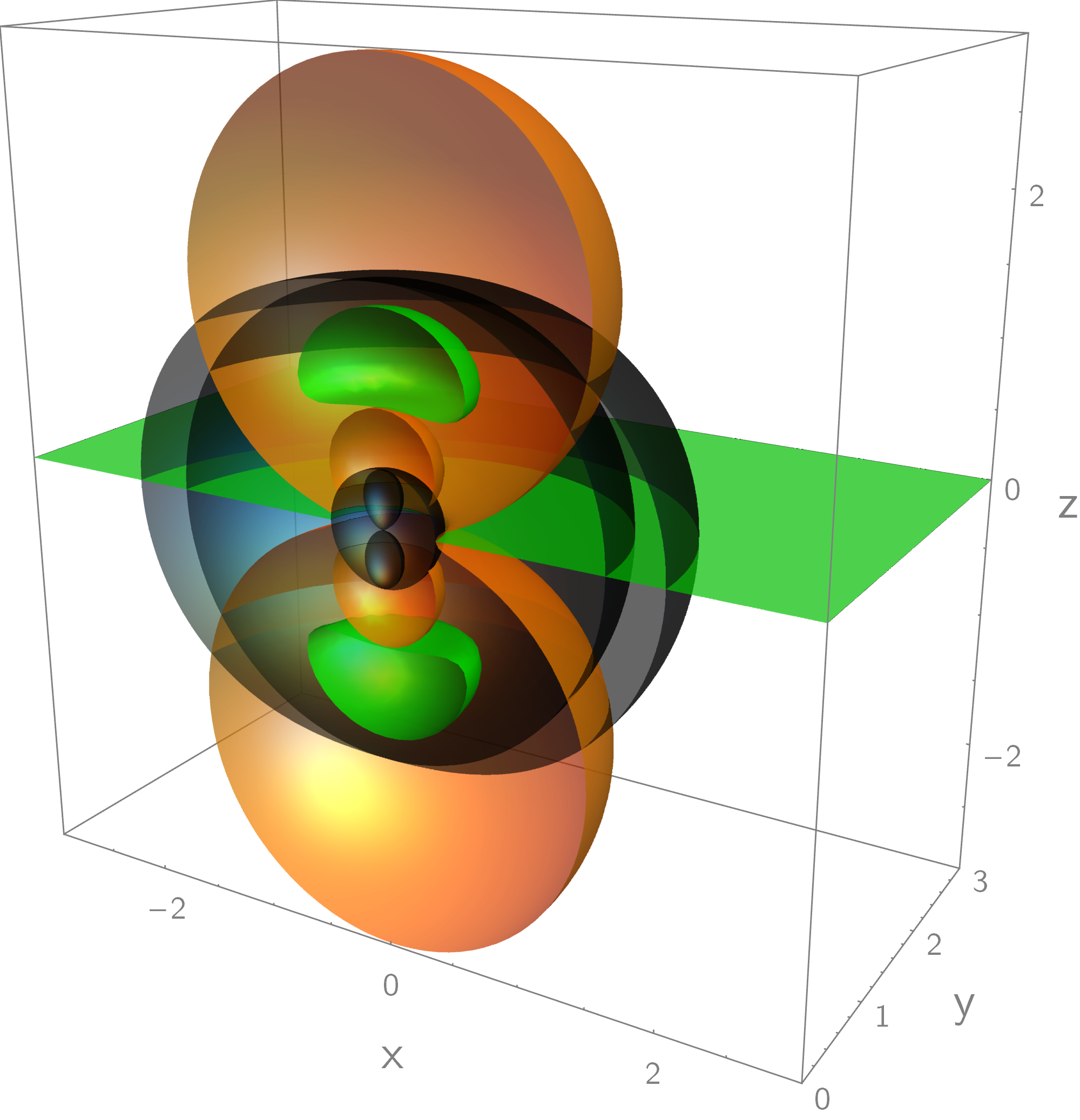} }}
    \qquad
    \subfloat[$m=1$, $a=0.99$]{{\includegraphics[width=0.38\textwidth]{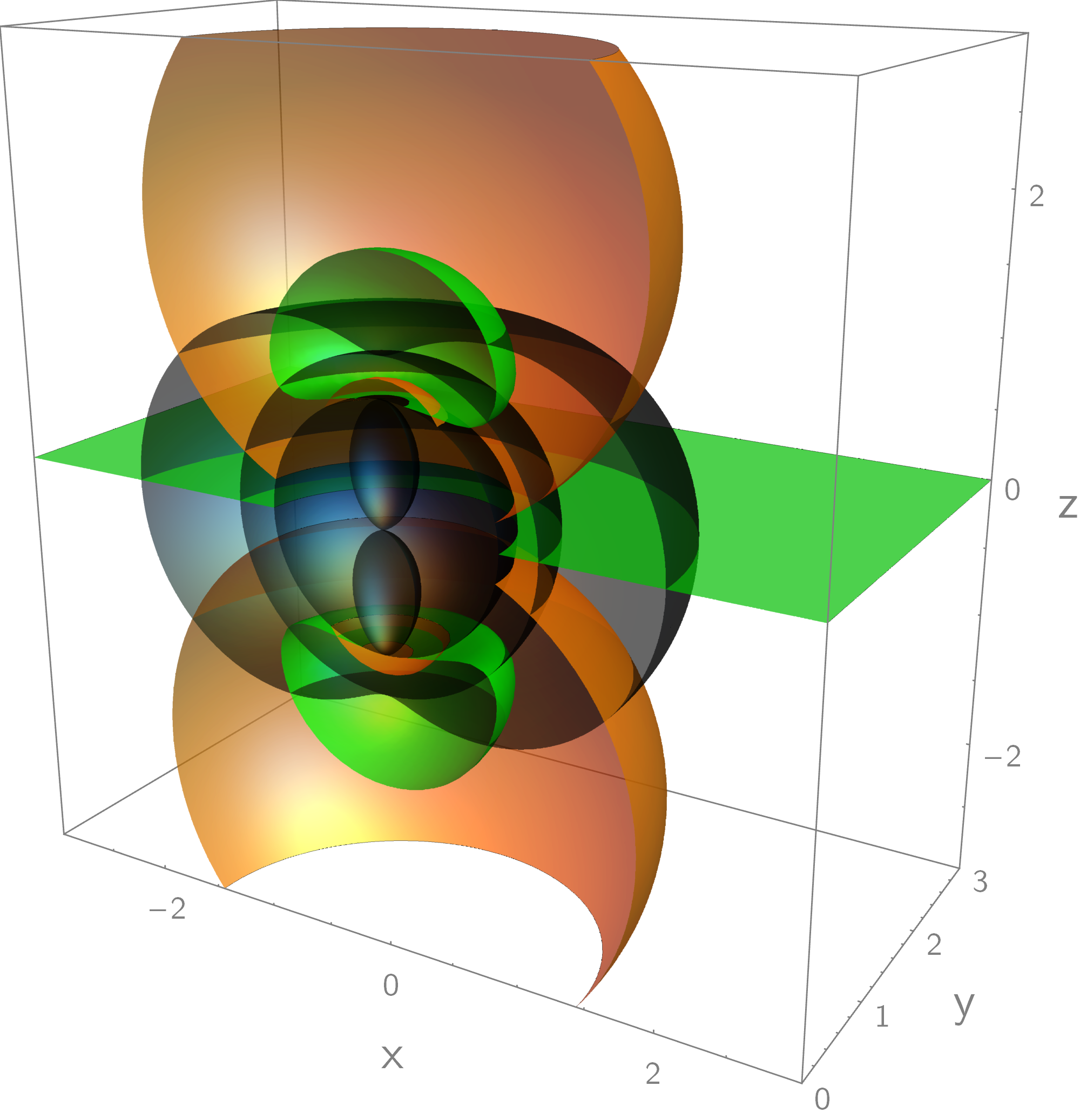} }}%
	\caption{(Colors online.) Surfaces of vanishing Kretschmann scalar (orange) and vanishing Pontryagin pseudoscalar (green) visualized for Kerr black holes of different parameters, and $r^2 =: x^2+y^2+z^2$. The inner and outer ergosphere, as well as the inner and outer horizon, are plotted in black. We excluded the interior of the inner horizon from the plots in order to improve visibility of the structures close to the horizon.}
    \label{fig:3d-zero-curvature}
\end{figure*}

Of course, one special value still exists: zero. Suppose we consider the dynamics of a non-minimally coupled matter field in the vicinity of a black hole. Under some assumptions on the Lagrangian, the non-minimally coupled curvature expressions can serve as an effective potential for the matter field. Therefore, in a WKB approximation where the field dynamics are fast compared to the gravitational dynamics, the field may condense in the minimum of its potential. Notably, in some circumstances, the expression of this minimum may have the structure of Eq.~\eqref{eq:invariant-curvature-surface}, see Fig.~\ref{fig:3d-zero-curvature} for a few zero-curvature surfaces around the Kerr black hole. Again, they may extend far outside the horizon.

\section{Projective surfaces and principal null directions}
Suppose we have a vector field at our disposal: then, we may consider scalar invariants formed by contractions of expressions in the curvature and that vector field:
\begin{align}
\left[f{}_{a_1\,\dots\,a_p}(C{}_{ijkl})\right] n{}^{a_1} \dots n{}^{a_p} = \text{const} .
\end{align}
An important class of vector fields, $n{}^i$, intrinsic to a given spacetime, are the principal null directions (PNDs) of the Weyl tensor, defined by $n{}^{[i}C{}^{j]mn[k} n{}^{l]} n{}_m n{}_n = 0$ with $n^2=0$ \cite{Stephani:2004}. The algebraic multiplicity of the corresponding eigenvalue problem then defines the Petrov type of the given spacetime at each point. The Kerr spacetime is of type D, and the PNDs are
\begin{align}
n{}^i_\pm = \left( \frac{r^2+a^2}{\Delta}, \pm 1, 0, \frac{1}{\Delta} \right), ~ n{}^\mu_\pm = \sqrt{\frac{\rho^2}{\Delta}}(1, \pm 1, 0, 0) .
\end{align}
Here, $n{}^\mu_\pm$ denotes the components of the PNDs with respect the coframe $\vartheta{}^\mu$. According to Eq.~(6) in Bel's work \cite{Bel:1962}, there is an equivalent way to formulate this eigenvalue problem for Petrov type D spacetimes (what Bel calls type IIb). For the Kerr solution it reads
\begin{align}
\begin{split}
C{}^\mu{}_\alpha{}^\nu{}_\beta n{}^\alpha_\pm n{}^\beta_\pm &= -2\mathbb{E}n{}^\mu_\pm n{}^\nu_\pm , \\
\left(\ast C\right){}^\mu{}_\alpha{}^\nu{}_\beta n{}^\alpha_\pm n{}^\beta_\pm &= +2\mathbb{B}n{}^\mu_\pm n{}^\nu_\pm .
\end{split}
\end{align}
It is straightforward to check that the above implies
\begin{align}
\begin{split}
B{}_{abcd} n{}^a_\pm n{}^b_\pm n{}^c_\pm n{}^d_\pm = 0 , \quad K{}_{abcd} n{}^a_\pm n{}^b_\pm n{}^c_\pm n{}^d_\pm = 0 . \label{eq:bel-and-kummer-surfaces}
\end{split}
\end{align}
Is the converse also true? Interestingly, inserting instead the general null vector $v{}^\mu = \left(\pm\sqrt{v_1^2 + v_2^2 + v_3^2}, v_1, v_2, v_3 \right)$ into the left-hand side of Eq.~\eqref{eq:bel-and-kummer-surfaces} implies
\begin{align}
m^2\left(v_2^2+v_3^2\right)^2 = 0 , \quad m^2 \mathbb{E} \left(v_2^2+v_3^2\right)^2 = 0 ,
\end{align}
respectively. For $m\not=0$ and $\mathbb{E}\not=0$, the unique solution is $v_2=v_3=0$ for any $v_1$. Hence either the Bel--Robinson surface or the Kummer surface imply the PNDs of the Kerr spacetime. Due to the algebraic nature of this proof, it seems plausible to us that this result may hold for general type D spacetimes. It remains to be seen whether these concepts can be generalized to different Petrov types.

\section{Conclusions}
Invariant curvature surfaces and projective surfaces seem to play an important role in the study of the Kerr geometry, both for experimental reasons (non-minimally coupled matter fields) as well as for theoretical considerations (Petrov classification). More work is necessary to extend our conclusions beyond the Kerr metric to general Petrov type D solutions and perhaps to other algebraically special spacetimes as well.

\begin{acknowledgments}
Discussions with Friedrich W.~Hehl (Cologne \& Missouri) are greatly appreciated. JB was supported by a Doctoral Recruitment Scholarship and the Golden Bell Jar Graduate Scholarship in Physics at the University of Alberta.
\end{acknowledgments}

\vspace{3cm}


\end{document}